# Quantum And Classical Dynamics Of Atoms In A Magneto-optical Lattice

Shohini Ghose, Paul M. Alsing and Ivan H. Deutsch

*Dept. of Physics and Astronomy, University of New Mexico, Albuquerque, NM, 87131*

Poul S. Jessen and David L. Haycock

*Optical Sciences Center, University of Arizona, Tucson, AZ 85721*

Tanmoy Bhattacharya, Salman Habib and Kurt Jacobs

*T-8, Theoretical Division, MS B285, Los Alamos National Laboratory, Los Alamos, NM 87545*

**Abstract**. The transport of ultra-cold atoms in magneto-optical potentials provides a clean setting in which to investigate the distinct predictions of classical versus quantum dynamics for a system with coupled degrees of freedom. In this system, entanglement at the quantum level and chaos at the classical level arise from the coupling between the atomic spin and its center-of-mass motion. Experiments, performed deep in the quantum regime, correspond to dynamic quantum tunneling. This nonclassical behavior is contrasted with the predictions for an initial phase space distribution produced in the experiment, but undergoing classical Hamiltonian flow. We study conditions under which the trapped atoms can be made to exhibit classical dynamics through the process of continuous measurement, which localizes the probability distribution to phase space trajectories, consistent with the uncertainty principle and quantum "back-action" noise. This method allows us to analytically and numerically identify the quantum-classical boundary.

## INTRODUCTION

Coherent control is one of the great challenges in contemporary physics with applications ranging from engineered chemical reactions [1] to electron transport in semiconductors [2-5] . The issues are richest in complex systems with multiple coupled degrees of freedom. At the quantum level, the coupling between the various subsystems can lead to highly entangled states with no classical description. Such entangled states play a major role in various areas of quantum information processing [6] . In the classical limit, systems with coupled degrees of freedom will exhibit nonlinear dynamics and chaotic motion. At the theoretical level, one seeks to better understand the border between the distinct predictions of classical and quantum dynamics, and to perhaps ultimately to control the system's behavior across this boundary.

Various studies of quantum systems with a chaotic classical limit have been carried out. While most have focused on static aspects such as wave function scars and energy spectrum statistics [7, 8] , some have investigated the dynamical features of these systems. The experiments of Raizen *et al.*[9, 10] have used ultra cold atoms trapped in standing waves of light (optical lattices) to explore the phenomenon of dynamical localization and the effect of the environment on a kicked rotor system. These experiments demonstrate that the atom/optical realization provides a very clean arena in which to study coherent quantum dynamics versus nonlinear classically chaotic motion.

We have explored a new system in which to study the rich quantum and classical dynamics associated with coupled systems – ultra-cold atoms trapped in far off resonance magneto-optical lattices. Like the kicked rotor, this system has several attractive feature for experiments: the potential can be modeled and designed with great flexibility, and state preparation, controlled unitary evolution and measurement can be performed efficiently through a combination of well established techniques of laser cooling, optical pumping and application of magnetic fields. Coherence times can be very long, and interaction with the environment (dissipation) can be introduced in a controlled manner. Finally, the ability to continuously measure the system enables us to explore the quantum-classical boundary and the emergence of chaos from quantum dynamics. In contrast to the kicked rotor where chaos is due to an externally applied time-dependent perturbation, in our system nonlinearity can arise from two coupled degrees of freedom, namely the atomic magnetic moment and its motion in the lattice. We can thus explore how the quantum-classical boundary is crossed as each of these subsystems varies from microscopic to macroscopic.

In this paper we review our recent experimental and theoretical studies of atomic transport in magneto-optical lattices. In section II we summarize the main features of this trap for alkali atoms. In the classical limit the coupling of spin and motion has the form of the motion of a magnetic moment moving in a spatially inhomogeneous magnetic field. This can lead to chaotic dynamics (section III). We describe the experimental setup and results in section IV. Comparison of the experimental data to quantum and classical theoretical predictions demonstrates the non-classical nature of the observed dynamics. We further show that the motion corresponds to dynamical tunneling through a potential barrier that depends on the internal state of the atom. In section V we explore the key question of measuring classical chaotic trajectories in this system. We study the effect of measurement back-action on our coupled quantum system and the conditions under which classical behavior is recovered. Finally, we end with a summary of our main results in section VI.

## ALKALI ATOMS IN A MAGNETO-OPTICAL LATTICE

Optical lattices are formed by the ac-Stark effect arising from the interaction of individual atoms with a standing wave created by a set of interfering laser beams. The physics of this system has been previously described in [11, 12] . Here we briefly summarize the main features. Our one-dimensional optical lattice is formed by two counterpropagating plane waves whose linear polarization vectors are misaligned by a relative angle $\Theta_L$. The resulting field can be decomposed into $\sigma+$ and $\sigma-$ standing waves whose nodes are separated by $\Theta_L/k$, where $k$ is the laser wave number. Consider for simplicity, an alkali atom whose ground state valence electron has spin-1/2. The resulting dipole interaction as a function of the atomic position, $z$, can be cast as the sum of a scalar part (independent of the atom's spin state) and a vector part which appears as an effective Zeeman interaction [11],

$$U(z,\vec{\mu}) = U_J(z) - \vec{\mu} \cdot \mathbf{B}_{eff}(z) \qquad (1)$$

Here $U_J(z) = 2U_0 \cos\Theta_L \cos(2kz)$, where $U_0$ is a constant depending on the atomic polarizability and field intensity. The effective magnetic field, $\mathbf{B}_{eff}(z) = B_x \mathbf{e}_x + B_{fict}(z)\mathbf{e}_z$, is the sum of an applied transverse magnetic field, $B_x$, plus a fictitious field associated with the ellipticity of the optical lattice laser polarization. For our geometry, $\mu_B \mathbf{B}_{fict} = -U_0 \sin\Theta_L \sin(2kz)\mathbf{e}_z$, where $\mu_B$ is the Bohr magneton. In actual experiments, we consider alkali atoms interacting with a 1-D lattice far detuned from the $nS_{1/2} \to nP_{1/2}$ resonance (the D2 line). In this far off resonance limit, the form of the potential (Eq. (1)) remains unchanged. The atomic magnetic moment in this case equals $\vec{\mu} = \hbar\gamma\mathbf{F} = -\mu_B\mathbf{F}/F$, where $\gamma$ is the gyromagnetic ratio and $\mathbf{F}$ is the total angular momentum vector of the atomic hyperfine ground state (including electron and nuclear spin) in units of $\hbar$. We consider here $^{133}$Cs, with $F=4$, the atom used in our current experiments [13]. The eigenvalues of the potential as a function of position result in nine adiabatic potentials (Fig. 1a), the lowest of which exhibits a lattice of double-wells.

The effective magnetic field causes Larmor precession of the atom's magnetic moment direction. Due to the coupling between the atomic position and its magnetic moment through the fictitious magnetic field, the spin precession is accompanied by motion of the wave packet between the double wells. This correlation between the internal state precession and center of mass motion leads to entangled spinor wave packets. The oscillation of the magnetization thus provides a meter through which we can detect the time dependent motion of the packet. Experiments of this type are described in more detail below.

## CHAOTIC DYNAMICS

The classical analog to the Hamiltonian associated with the far off resonance magneto-optical potential corresponds to the motion of a massive particle with a magnetic moment moving in a combination of a scalar potential (independent of the moment) plus a spatially inhomogeneous magnetic field [14]. The classical equations of motion then have the form

$$\dot{z} = p/m, \quad \dot{p} = -\frac{d}{dz}\left(U_0(z) - \vec{\mu} \cdot \mathbf{B}_{eff}(z)\right), \quad \dot{\vec{\mu}} = \gamma\left(\vec{\mu} \times \mathbf{B}_{eff}(z)\right), \qquad (2)$$

where $\mathbf{B}_{eff}(z) = B_{fict}(z)\mathbf{e}_z + B_x \mathbf{e}_x$, with fictitious field as given in the previous section. These equations follow from the Heisenberg equations of motion, replacing the quantum operators by their expectation values and neglecting any correlations in the operator products. The effective phase space is four dimensional with two external and two internal variables. The Hamiltonian is in general non-integrable but can be made integrable under two simple physical circumstances: the case in which there is no transverse magnetic field, $B_x$, [15], and the case of a sufficiently large transverse field so that the motion is adiabatic [16]. We study each regime separately using a set of canonical action-angle variables.

When there is no transverse magnetic field $(B_x = 0)$, $n_z \equiv \mu_z/|\vec{\mu}|$ becomes an additional constant of motion. This results in an integrable Hamiltonian that is identical to that of a simple pendulum

$$H_0 = \frac{p^2}{2m} + C\cos(2kz + D), \qquad (3)$$

$$C = U_0\sqrt{4\cos^2\Theta_L + n_z^2\sin^2\Theta_L}, \quad D = \arctan(n_z\tan\Theta_L/2). \qquad (3a)$$

whose amplitude and phase depend on the constant z-projection of the atomic moment [17]. The action-angle variables for a pendulum, $(J,\psi)_-$ are well known to be functions of the complete elliptic integrals [18]. For energies close to the bottom of the sinusoidal potential, we can expand the elliptic integrals in a power series, keeping only the first few terms. This enables us to express $H_0$ as a function of the motional actions $J$ and internal action $\mu_z/\gamma$. The precession frequencies of the corresponding angle variables $\psi$ and $\chi$ can then be computed from Hamilton's equations to be,

$$\omega_1 = \dot\psi = \frac{\partial H_0}{\partial J} = \frac{\pi}{2}\frac{\omega_0}{K(\kappa)}, \quad \omega_2 = \dot\chi = \gamma\frac{\partial H_0}{\partial \mu_z} = \frac{\partial C}{\partial n_z}\frac{\gamma H_0}{\mu_B C}, \qquad (4)$$

where $\omega_0 = \sqrt{4k^2|C|/m}$ is the oscillation frequency for a harmonic approximation to the sinusoidal potential, $2\kappa^2 = 1 + H_0/|C|$, and $K(\kappa)$ is the complete elliptic integral of the first kind. The frequency $\omega_1$ represents oscillation of the center of mass in the sinusoidal potential. By moving to a frame that oscillates with the atom, the time dependence in the effective magnetic field is removed, resulting in a *constant* precession frequency $\omega_2$ about the z-axis. The precession angle in this frame is $\chi$. The addition of a transverse magnetic field as a small perturbation to this integrable Hamiltonian couples the oscillations of the two angles, giving rise to nonlinear resonances. The primary resonances occur when the ratio of the unperturbed frequencies is a rational number, and can be calculated for our system using Eqs. (4).

Current experiments operate in the adiabatic regime where the applied transverse field is large and thus cannot be treated as a perturbation as outlined above. We therefore analyze the regime in which the non-adiabatic coupling can be treated as a perturbation. The integrable adiabatic Hamiltonian is obtained by setting the angle $\alpha$ between $\vec{\mu}$ and $\mathbf{B}_{eff}$ to be a constant, so that

$$H_0 = p^2/2m + U_J(z) + \mu_B |\mathbf{B}_{eff}(z)| \cos\alpha. \tag{5}$$

When $\alpha = 0$ we obtain the lowest adiabatic double-well potential (Fig. 1a). Other fixed values of $\alpha$ correspond to other adiabatic surfaces. The component of $\vec{\mu}$ along the direction of the magnetic field is now a constant of motion and serves as our new action variable. The other action of the system is obtained in the standard way by integrating the momentum over a closed orbit in the double well for a given energy and $\alpha$. The precession frequencies $\omega_1$ and $\omega_2$, of the conjugate angle variables correspond respectively to the oscillation of the center of mass in the adiabatic double-well potential and precession of the magnetic moment about the local magnetic field direction in a frame oscillating with the atom as described previously. Unlike the previous case however, we cannot obtain analytical expressions for the frequencies and must resort to computing them numerically. Figure 1(b) shows a typical surface of section for motion in this regime. The primary resonance at $n_z = 0.38$ and $\phi = 0$ corresponds to a ratio of the unperturbed adiabatic frequencies of $\omega_2/\omega_1 = 4$. The nonadiabatic perturbative coupling is strong enough to cause the previously stable primary resonance at $n_z = 0.8$ to bifurcate, and secondary resonances to appear around the points $n_z = 0.38$ and $n_z = -0.85$. The secondary resonances result from coupling between the motion around the primary islands to the unperturbed periodic motion. As the energy is increased, the primary resonances eventually disappear and global chaos sets in.

## QUANTUM VS. CLASSICAL DYNAMICS

We have recently performed experiments to observe quantum transport of atoms in the magneto-optical lattice in a mesoscopic regime [13] . We prepare a sample of about $10^6$ Cs atoms in a well-defined initial quantum state, say $|\psi_L\rangle$, and follow their subsequent quantum coherent evolution. We start out by laser cooling in a standard magneto-optical trap/3D molasses to a temperature $\sim 4\mu K$ and a Gaussian density distribution of ~200 _m RMS radius, followed by further cooling in a near-resonance 1D lin-$\theta$-lin lattice. The atoms are then transferred to a far off resonance (detuned 3000 linewidths from resonance) 1D lin-$\theta$-lin lattice and optically pumped to $|m_F = 4\rangle$. In order to select the motional ground state in the corresponding potential, the depth of the lattice is lowered and it is accelerated at $300\,m/s^2$ for 1.5ms so that atoms outside the ground band can escape. Optical pumping and state selection is done in the presence of a large longitudinal external field $B_z$ (-55mG) to lift degeneracies between the optical potentials and prevent precession of the magnetic moment. This state selection procedure prepares the atoms with roughly 90% population in the target state. Once this is achieved, we increase the lattice depth, change the lattice acceleration to free-fall, ramp up the transverse field $B_x$ and finally ramp $B_z$ to zero. By performing this sequence slowly enough, we adiabatically connect the ground state in the $m_F = 4$ potential to the left-localized state of the lowest optical double well.

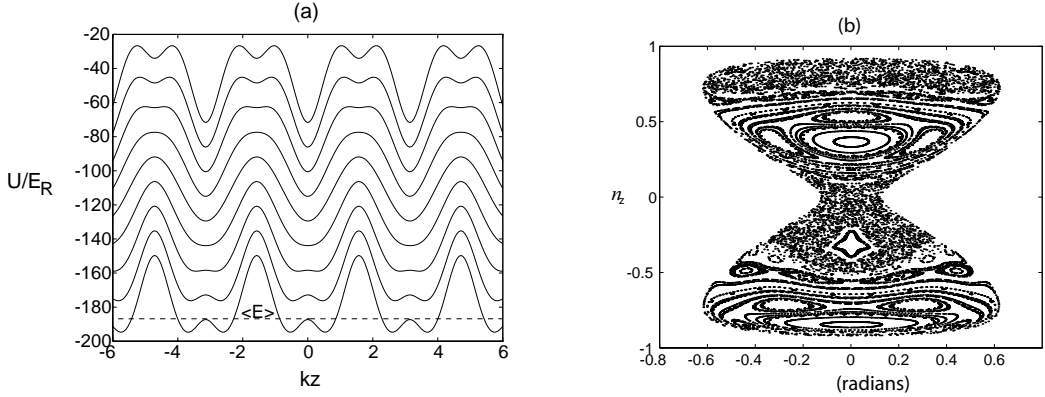

**FIGURE 1**. (a) Adiabatic potentials for Cesium atoms in an optical lattice with an additional external transverse magnetic field. The mean energy of the state prepared in experiments [13] is just greater than the lowest adiabatic potential barrier energy (horizontal line). The Poincaré surface of section in (b), for $p = 0$ and $dp/dt > 0$ using the parameters of the experiment [13], with $E = -186.8E_R$ ($E_R/h = 2kHz$), shows the effects of the non-adiabatic perturbation term, which makes the classical equations (Eq.(2)) non-integrable.

The motion of the atoms in the double well is measured indirectly by monitoring the time-dependent magnetization of the atomic ensemble. A Stern-Gerlach measurement is performed by releasing the atoms from the lattice, quickly applying a bias field $B_z$ to keep the quantization axis well defined, and letting the atoms fall to a probe beam a few cm below in the presence of a magnetic field gradient of 13G/cm. The magnetic populations can then be extracted from the separate arrival time distributions for different magnetic sublevels.

In order to understand the classical vs. quantum nature of the atomic motion, we compare the experimental data to the predictions of a fully classical calculation as well as a quantum bandstructure calculation [14]. Given the initial state, we compute the joint Husimi or "$Q$" quasiprobabiluty distribution over both external and internal phase space by employing the familiar motional coherent states as well as the spin coherent states [19]. This initial distribution is evolved both classically and quantum mechanically and compared to the experimental results. The contrast between classical and quantum dynamics is clearly shown in Figure 2(a). Due to the correlation between the atomic magnetic moment and its motion in the well, an oscillation of the mean magnetization between positive and negative values implies motion of the atom from one minimum of the double well to the other. Classical dynamics thus predicts that the mean of the distribution remains localized on one side of the double well. In contrast, the experimental data shows an oscillation between positive and negative values at a frequency well predicted by the quantum model.

A closer look at the reduced classical distribution in the phase space of position and momentum, obtained by tracing over the magnetic moment direction, shows that a part of the distribution does move between wells, but the peak remains localized in one well (Fig. 2b). This implies that transport between the wells is not classically forbidden, but is unlikely for this distribution of initial conditions. This is due to the fact that the classical description of the state corresponding to the $Q$-function involves a *distribution* of energies. High energy tails of this distribution can classically hop between the left and right wells. However, the experimentally observed oscillations of the mean atomic magnetization are much better described by quantum evolution indicating a *nonclassical* motion of the atom between the double wells. This is not surprising given the fact that for the given experimental parameters, the actions of the system are on the order of $\hbar$ (External center-of-mass action, $I_0 \approx 10\hbar$ and spin $J = 4\hbar$). The only discrepancy with the quantum model is the experimentally observed decay of the oscillations. The probable cause is an estimated ~5% variation of the lattice beam intensities, which is consistent with the observed dephasing times. We estimate the timescale for decoherence due to photon scattering to be of order ~1ms, which is too slow to account for the observed damping. A next generation of the experiment is now underway, in which we hope to increase the dephasing time by an order of magnitude by better control over lattice beam and magnetic field inhomogeneities.

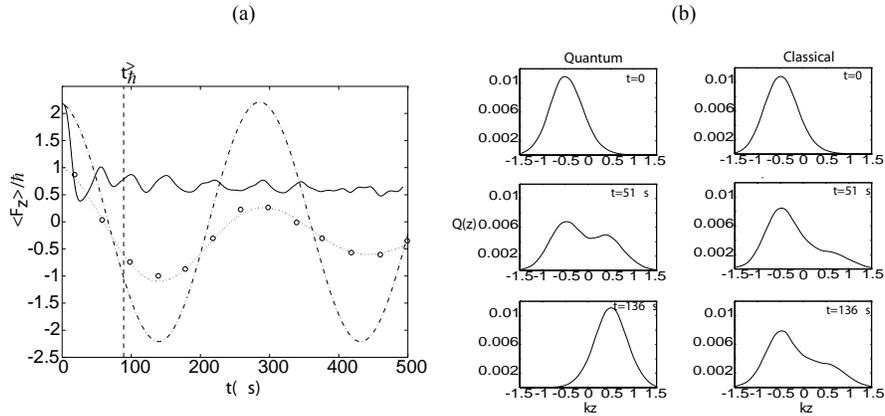

**FIGURE 2**. (a) Predictions of mean magnetization dynamics. Ideal quantum theory: two-level Rabi-flopping (dashed-dotted); Ideal classical theory: localized at positive $\langle F_z \rangle$ (solid); Experimental: (circles) with a damped sinusoid fit. (b) Reduced Q–distribution in position $Q(z)$, at different times in the quantum versus classical evolution. The quantum distribution oscillates between wells while the classical distribution remains mostly on the left side, with a portion equilibrating between the wells.

We can ask whether the observed non-classical motion can be interpreted as 'quantum tunneling'. For particles with more than one degree of freedom, the identification of tunneling behavior can become ambiguous since the total energy does not uniquely define the particle's classical trajectory. In particular, for the magneto-optical lattice at hand, the potential energy depends not only on the position of the atom, but also on its internal state in a correlated way. The initial state prepared in the experiment mostly populates the lowest adiabatic potential, but at times corresponding to a Schrödinger cat-like superposition in the two wells, there is a small component in the second lowest potential due to the nonadiabatic coupling. Whereas the mean energy of the population in the lowest adiabatic potential is higher than the corresponding double well barrier, the barrier of the next adiabatic potential is much higher than the mean energy of the small population in that potential. The oscillation of the population from right to left well in this adiabatic potential corresponds to tunneling through a classically forbidden regime. The atom therefore sees a population-weighted average of the two lowest adiabatic potential barriers [14]. The non-adiabatic transitions of the internal state cause the tunneling barrier to be dynamical in nature.

## QUANTUM-CLASSICAL TRANSITION

Given the disparity between the classical and quantum phase space dynamics, one can ask under what circumstances classical dynamics is recovered? In recent years, it has been widely appreciated that emergent classical behavior can arise when the quantum system is weakly coupled to an environment. Decoherence resulting from tracing over the environment can suppress quantum interference and, in many circumstances, this can lead to an effectively classical evolution of a phase space distribution function [20, 21]. While decoherence can explain classical behavior for *mean values* of observables, it does not succeed in extracting localized "trajectories" from the quantum dynamics. Such trajectories are useful for quantifying the existence of chaos both theoretically and in experiments through the quantitative measure of the Lyapunov exponents. One can recover trajectories from the quantum dynamics when the environment is taken to be a meter that is continuously monitored, leading to an evolution of the system density operator conditioned on the measurement [22]. If one averages over all possible measurement results, the description reverts to that at the level of phase space distributions.

Continuous measurement provides information about the state of the system and thus localizes it in phase space. These localized trajectories have added quantum noise, however, due to quantum measurement backaction. Therefore, in order to recover the desired classical trajectories, the system must be in a regime where the measurement causes strong localization but weak noise [23].

We have studied both numerically and analytically the conditions for recovering classical behavior in our system given a continuous measurement of the atomic position [24]. We take as our model Hamiltonian,

$$H = \frac{p^2}{2m} + \frac{1}{2}m\omega^2 z^2 + bz\frac{J_z}{J} + c\frac{J_x}{J} \qquad (6)$$

which is a harmonic approximation to the magneto-optical lattice potential at a single lattice site. The evolution of the system conditioned on a record of the atoms mean position, $\langle z \rangle + (8k)^{-1/2} dW/dt$, is studied using a stochastic Schrödinger equation [25, 26],

$$d|\tilde{\psi}\rangle = \left\{ \frac{1}{i\hbar} H dt - kz^2 dt + \left(4k\langle z \rangle dt + \sqrt{2k} dW\right) z \right\} |\psi\rangle \qquad (7)$$

where the tilde denotes an unnormalized quantum state, $k$ is the "measurement strength", and $dW$ describes a Wiener noise process. Here we have assumed perfect measurement efficiency. We can numerically evolve this equation using a Milstein algorithm [27] for the stochastic term. We pick as our initial condition a product of minimum uncertainty states (coherent states) in position and spin, and compare to the classical trajectories centered at the same initial points; we choose the initial spin coherent state in the *x*-direction though any direction in the *x*-*y* plane would be equivalent. We fix $b = -m\omega^2 \Delta z / J$, with $\Delta z \approx 15 z_g$ where $z_g$ is the ground state rms width of the wells. Previous work [23] has established a window of measurement strengths for sufficiently large external (center of mass) actions, which satisfy the dual desire for strong localization and weak measurement back-action. We build upon that result here, choosing $I \approx 1000\hbar$. This puts us in a semi-classical regime where the external degree of freedom is effectively classical but the quantum nature of internal dynamics can be still be important.

We first consider the dynamics of the smallest spin system, *J*=1/2 in the integrable regime ($c = 0$). We see in Fig. 3(a) that the quantum trajectory quickly diverges from the classical trajectory. This can be understood by studying the effect of the position measurement on the spin subsystem. The initial spin state pointing in the *x*-direction is an equal superposition of spin-up and spin-down states, which move along the wells centered at $z_\uparrow = -\Delta z$ and $z_\downarrow = \Delta z$ respectively (Fig. 4 a,c). The two spin components of the initial spatially localized wave packet spatially thus separate into a left and a right wavepacket, so that the total wave function evolves into an entangled Bell-like state, $|\psi(t)\rangle = |\phi_{left}\rangle|\uparrow\rangle + |\phi_{right}\rangle|\downarrow\rangle$, with $\langle \phi_{left} | \phi_{right} \rangle \neq 1$. This splitting of the wave packet causes an initial rapid increase of the position variance (outer solid curves in Fig. 3a). When the left and right components of the state become spatially resolvable beyond measurement errors, the position measurement, acting as a meter for spin, collapses the wave function into either the left potential (spin-up state) or the right potential (spin-down state). This contrasts the fully classical dynamics, which predicts oscillation about the origin.

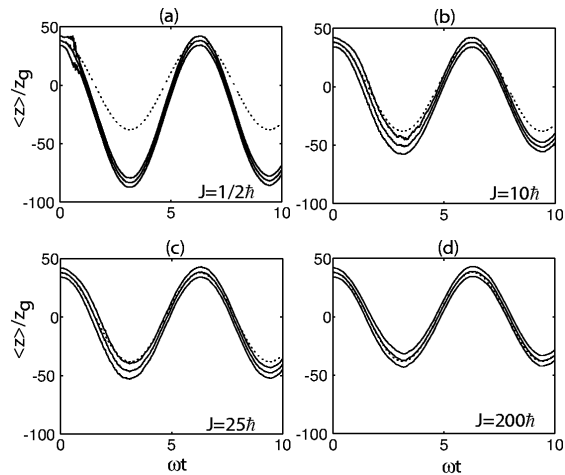

**FIGURE 3:** Mean position of the measured system (solid) in a single quantum trajectory for different values of spin with $\Delta z \approx 15 z_g, I \approx 1000\hbar, k = \omega/2 z_g^2$ Outer solid curves show the variance of the wave function. As J gets larger the mean position evolution approaches the classical (dotted) trajectory.

When the magnitude of the spin, *J*, becomes much larger than 1/2, an initial spin coherent state in the *x*-direction is no longer a superposition solely of spin up and down states in the *z*-basis, but rather, a distribution over all 2*J*+1 $M_J$ states, centered around $M_J = 0$. Just as in the spin 1/2 case, an initially localized wave function will spread out in space as the different spinor components move along the different potentials centered at $z_{M_J} = -(M_J / J)\Delta z$, (Fig 4. b,d). However, as *J* becomes larger, the population distribution becomes more peaked at the $M_J = 0$ state, so that most of the population moves along the potential centered at $z_0 = 0$, which corresponds to the classical potential. The measurement is thus more likely to localize the atom in this classical potential and damp out the tails of the wave function that spread out over the outermost potentials. The key point is that the effective spin measurement in the $J_z$ basis is no longer strongly projective, and therefore the weak noise condition can be met along with the strong localization condition.

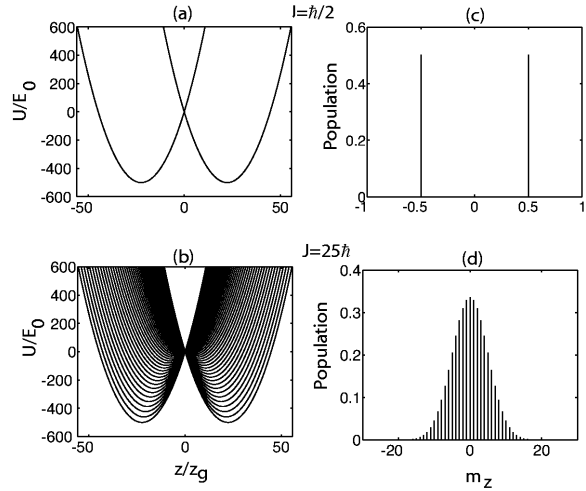

**FIGURE 4**. The spin up and down components of a spin 1/2 wave function move along 2 different potential wells (a). For J >>1/2, the spin components of the wave function evolve along 2J+1 different potentials (b). Histograms for the population in each mz state (c,d) show that as J gets larger the population becomes peaked around the $m_z$=0 state (d). The position is thus more likely to localize the wave function in the central (classical) potential well.

The scale of *J* for which the weak noise and strong localization conditions are satisfied can be analytically determined by following the approach in [23]. The stochastic equations of motion for the mean position and momentum follow from Eq. (7),

$$d\langle z \rangle = \frac{\langle p \rangle}{m} dt + \sqrt{8k} C_{zz} dW$$
$$d\langle p \rangle = -m\omega^2 \langle z \rangle dt - b\langle J_z \rangle dt + \sqrt{8k} C_{zp} dW \qquad (8)$$
$$d\langle J_z \rangle = \sqrt{8k} C_{zJ_z} dW$$

where $C_{ab} = (\langle ab \rangle + \langle ba \rangle)/2 - \langle a \rangle \langle b \rangle$ are the symmetrized covariances. The small noise and strong localization conditions applied to these equations can be combined into the condition that the covariance matrix in the ordered basis $\{z,p,J_z\}$ remain small at all times relative to the allowed phase space of the dynamics. We have ignored the $x$ and $y$ components of the angular momentum since we are interested in measuring the position of the particle, which depends only on $J_z$. Furthermore, we neglect third and higher cumulants in the evolution of the covariance matrix $C$, since these remain small for large $J$. Under this approximation, the covariance matrix evolves according to a matrix Riccati equation, $\dot{C}(t) = \alpha + \beta R(t) + R(t)\beta^T + R(t)\gamma R(t)$ [24]. This equation has an analytical solution since the matrices $\alpha, \beta$ and $\gamma$ are time independent [28]. However the analytical solutions for the second cumulants are not simple functions of the system parameters. We resort to numerically finding the bounds on the analytical solutions $C(t)$. Our studies indicate that the bounds on $C(t)$ become small relative to the allowed phase space, only when $J$ is much larger than $\hbar$, as expected.

Our numerical and analytical results show that classical dynamics is only recovered in this coupled system when the actions of both subsystems become large relative to $\hbar$. When one subsystem lies in the quantum regime, even a weak measurement of the classical subsystem eventually results in a strongly projective measurement of the quantum subsystem, thus preventing the recovery of classical behavior. Preliminary numerical studies indicate that the same condition holds true in the chaotic regime. When the spin is large enough relative to $\hbar$, the measured trajectory recovers the mixed phase-space of the classical system. However, the analytical solutions of the corresponding Riccati equation for the covariance matrix are much more difficult to obtain since the matrix $\gamma$ is now time dependent. Work is in progress to find approximate bounds on the covariance matrix and to recover the classical Lyapunov exponent from the measured trajectories.

## SUMMARY

Atoms in optical lattices provide a clean testbed in which to study the rich dynamics of coupled systems. Our experiments employ Cesium atoms in a one-dimensional, far-off-resonance, magneto-optical lattice. Preparation of atoms in localized, pure quantum states in the double-potential wells is achieved through a combination of laser-cooling, optical pumping, and state-selection. Because of the correlation between the atomic position and internal state, precession of the atomic spin is accompanied by motion of the center-of-mass wave packet. The entanglement between atomic spin and center of mass motion provides a meter through which dynamical tunneling and other transport phenomena can be directly observed at the microscopic level. We have carried out a detailed study of the classically chaotic dynamics of our atom-lattice system, including a direct comparison between classical predictions and quantum theory/experiment. Our results underscore the profoundly non-classical nature of the observed tunneling Rabi oscillations. We find that under appropriate conditions the atom can "tunnel" through a dynamical energy barrier.

The manifestations of chaos in quantum mechanics and the emergent complexity at the classical level continues to be a problem of fundamental interest. Our numerical and analytical results show that classical chaotic trajectories can be recovered through continuous measurement of this system only when both the external and internal actions are large relative to $\hbar$. In future work we hope to continuously measure the spin of the system via Faraday rotation spectroscopy [29, 30] and develop more tools to explore the quantum-classical boundary.

# ACKNOWLEDGEMENTS

SG, PMA and IHD were supported by NSF Grant No. PHY-009569. PSJ and DLH were supported by grants NSF PHY0099582, ARO DAAD19-00-1-0375 and JSOP DAAD19-00-1-0359.